\documentstyle[12pt,epsf]{article}

\newcommand{\eq}[1]{eq.(\ref{#1})}

\def\be{\begin{equation}}
\def\ee{\end{equation}}

\def\Im{\mbox{Im}}
\def\Re{\mbox{Re}}

\def\half{{1 \over 2}}
\def\L{\Lambda}
\def\le{\lambda_{\mbox{eff}}}
\begin{document}
\begin{flushright}
PURD-TH-94-03 (rev.) \\
March 1994  \\
hep-ph/9401318
\end{flushright}
\vspace{0.5in}
\begin{center}
{\Large
Semiclassical transition probabilities for interacting oscillators} \\
\vspace{0.4in}
{\large S.Yu. Khlebnikov}
\\
\vspace{0.1in}
{\it Department of Physics, Purdue University,
West Lafayette, IN 47907, USA} \\
\vspace{0.5in}
{\bf Abstract} \\
\end{center}
Semiclassical transition probabilities characterize transfer of energy
between "hard" and "soft" modes in various physical systems. We establish
the boundary problem for singular euclidean solutions used to calculate
such probabilities. Solutions are found numerically for a system of two
interacting quartic oscillators. In the double-well case, we find
numerical evidence that certain regular {\em minkowskian} trajectories
have approximate stopping points or, equivalently, are approximately
periodic. This property leads to estimates of tunneling excitation
probabilities in that system and suggests that similar estimates may be
possible in other systems with tunneling.
\newpage
\section{Introduction}
Very little is known at present about semiclassical transition probabilities
in systems with more than one degree of freedom.
Let $x$ denote one of (possibly many) coordinate variables and
$|n\rangle$ be the $n$-th energy eigenstate. Then, a typical semiclassical
matrix element is $\langle n_2|x|n_1\rangle$ for $|n_2-n_1|\gg 1$.
In what follows we always assume $n_2>n_1$.
This matrix element appears  when one calculates the probability of
exciting a system originally in state $|n_1\rangle$ to state
$|n_2\rangle$ by applying a high-frequency external force proportional
to $x$. Similar matrix elements appear in other problems where there is
a transfer of energy between "hard" and "soft" modes.
The states $|n\rangle$ then refer to the "soft" subsystem.
In elementary particle
physics, a typical problem of this sort is calculation of cross sections
for production of many low-momentum particles in collisions of two
high-momentum ones. Electroweak baryon number non-conservation at high
energies is an example that was extensively discussed recently
(for most recent reviews see ref.\cite{reviews}) and there are other
interesting cases as well \cite{other}.

In a system with one degree of freedom, semiclassical transition probabilities
can be calculated by the method described in the Landau-Lifshitz textbook
\cite{LL}.
A natural generalization of this method to many degrees of freedom is
achieved by reinterpreting it in terms of singular classical solutions
\cite{IP}. In ref.\cite{mine} we have found some approximate
singular classical solutions
for self-interacting field theories in four dimensions. We have shown how
they can be used to
reproduce the factorial asymptotics of cross sections for multiparticle
production at relatively low energies and how the crossover to
a high-energy regime may occur.
This approach was pursued further in ref.\cite{DP}.

There are several problems preventing
a full-scale application of the method of singular solutions to field
theory. (i)  There is no formal derivation of the expression for transition
probability used in \cite{IP,mine}; consequently, some interpretation
problems arise as to exactly what quantity is being calculated.
(ii) No {\em exact} singular solutions of the required form are known in
systems with more than one degree of freedom (except when a reduction to
one degree of freedom occurs) and it is not clear what are
the existence and uniqueness conditions for such solutions.
(iii) In general, one expects that exact singular solutions are accessible
only numerically, and the appropriate numerical procedure needs to be
formulated.

The present work is
an attempt to get at least partial resolution of these problems
using the simplest non-trivial case of two interacting non-linear
oscillators as an example.  In Sect.2 we describe a derivation of
the double functional integral representation
for the following transition probability
\be
W(E_1,E_2)= \sum_{a,b} \Bigl|
\langle b| {\cal P}_{E_2} x {\cal P}_{E_1} |a\rangle \Bigr|^2 \; ,
\label{W}
\ee
where the sum is over a complete system of states
and  ${\cal P}_E$ is the projector onto the subspace of a given energy
$E$. We then show that in the semiclassical approximation,
the transition probability (\ref{W}) is saturated
by a singular solution on a certain contour in the complex time
plane. We will comment specifically on the appearance of time
dependence in the originally time-independent problem (\ref{W}).
In Sect.3 we present the results of a numerical search for
singular classical solutions in a system of two interacting quartic
oscillators.
We do not attempt a rigorous analytic resolution of the problem (ii) above
but rather present
numerical evidence for the existence and uniqueness
of singular solutions under specified boundary conditions.
The results of Sect.3 refer to the case of a single-well potential.
The double-well case, which is closer to the problem of electroweak
baryon number violation, is considered in Sect.4. We present there
numerical evidence that the {\em minkowskian} solution that starts with
zero velocity at a turning point of a periodic instanton
(which is a periodic {\em euclidean} solution \cite{period}) has
an approximate stopping point or, in other words,
is approximately periodic. As we illustrate in Sect.4, this property
leads to straightforward numerical estimates of tunneling excitation
probabilities in that double-well system. Similar estimates may be
possible in other systems with tunneling, maybe even in field theories.
Concluding Sect.5 contains a discussion of the results.

\section{Semiclassical formula for transition probability}
The projector ${\cal P}_E$ appearing in (\ref{W}) has a simple representation,
given in \cite{period}, only when the system reduces to a collection of
non-interacting harmonic oscillators. In field theory, this usually
happens at large positive and negative times when an initial configuration
dissociates into free particles. A system of a few degrees
of freedom, however, typically never gets
out of the non-linear regime.  To be able to use the representation
of ref.\cite{period} for ${\cal P}_E$ in the case of two interacting
oscillators, we need to approach the original problem in the following
way. Two interacting oscillators represent a particle confined by a potential
to some region on a two-dimensional plane. Let us couple this particle
to electromagnetic field by giving it some small charge. Then, an excited
state of the particle will eventually decay into photons, for which our
formula for the projector is applicable. For a highly excited state,
this decay can be viewed as an almost classical electromagnetic radiation by
the oscillating particle.
So, the coupling to electromagnetic field does not
spoil the semiclassical nature of our problem. Moreover, because the radiation
takes place in real time, the action on the corresponding part
of the solution does not contribute
to the exponential factor in the probability (\ref{W}). Thus, making
the electromagnetic coupling arbitrarily small, we make the probability
arbitrarily close to that in the system without electromagnetic coupling,
even though at large times the two systems look entirely different.

With this modification in mind, we can use the formula for ${\cal P}_E$
at sufficiently large times. On the other hand, the states in \eq{W}
are taken at some fixed moment of time $t=t_0$.
To proceed, we should define the state ${\cal P}_{E_1} |a\rangle$
at $t=-\infty$ and the state
${\cal P}_{E_2} |b\rangle$ at $t=\infty$, where we know how to do that,
then evolve these states to $t=t_0$ and use the result in \eq{W}.
As usual, it is convenient to define initial and final states in
the interaction representation. Then, eq.(\ref{W}) takes the form
\be
W(E_1,E_2)= \sum_b \Bigl| \langle b (\infty)|
{\cal P}_{E_2} S(\infty,t_0) x S(t_0,-\infty)
{\cal P}_{E_1} |a(-\infty)\rangle \Bigr|^2 \; ,
\label{W1}
\ee
where $S(\infty,t_0)$ is the "half of the $S$-matrix" operating from
$t=t_0$ to $t=\infty$. Thus, in order to use
the known expression for the projector we introduce time dependence
in the originally time-independent problem.
This may seem to be an additional complication but in fact it is just
the opposite. Instead of calculating difficult overlap integrals of
wave functions, we find the exponential suppression that is present in
$W(E_1,E_2)$ simply from the imaginary part of the action of a classical
solution.

In the leading semiclassical approximation, we are interested only
in the exponential factor in the probability, so the operator $x$ in \eq{W1}
is inessential. (Of course, without this operator the {\em pre-exponent} will
vanish.) The exponents of the two halves of the $S$-matrix in (\ref{W1})
add up to that of the whole $S$-matrix.
We can now proceed in a rather close analogy with the calculation of
ref.\cite{period}. The main difference with that calculation is in the
contour in the complex time plane, on which a classical solution is to
be found. When we consider a transition probability rather
than a transition amplitude, we actually need a contour that consists of
two parts symmetric with respect to the real axis. In the
present case, we use the contour shown in Fig.1 together with its reflection
into the lower half-plane.
\setlength{\unitlength}{1mm}
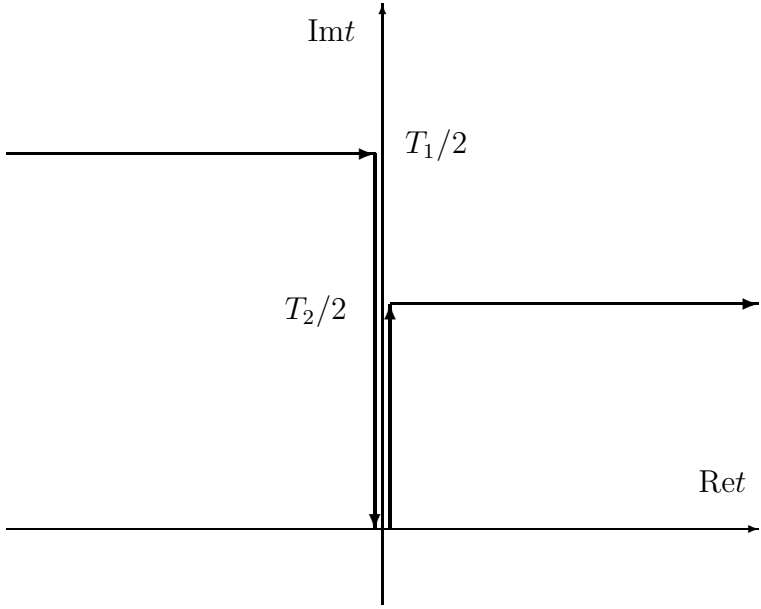
\begin{figure}
\begin{picture}(100,80)(-50,-10)
\put(-50,0){\vector(1,0){100}}
\put(0,-10){\vector(0,1){80}}
\thicklines
\put(-50,50){\vector(1,0){49}}
\put(-1,50){\vector(0,-1){50}}
\put(1,0){\vector(0,1){30}}
\put(1,30){\vector(1,0){49}}
\thinlines
\put(42,5){$\Re t$}
\put(-10,65){$\Im t$}
\put(3,50){$T_1/2$}
\put(-13,28){$T_2/2$}
\end{picture}
\caption{Contour in the complex time plane, on which
we look for a real classical solution.}
\end{figure}
Unlike the contour used in \cite{period},
the contour of Fig.1 has both euclidean and anti-euclidean parts,
in addition to minkowskian ones.
A classical solution along the contour shown in Fig.1
describes a transition from a state of energy $E_1$ at
$t=-\infty + i T_1/2$ to a state of energy $E_2$ at
$t=+\infty + i T_2/2$. This solution is necessarily singular because
the initial and final states have different energies. By convention,
the singularity is positioned at $t=0$.

It is important to realize how crucial is the fact that we consider
a total probability and therefore sum over $|a\rangle$
and $|b\rangle$ in \eq{W1}. If we were to consider probability
of transition between certain $|a\rangle$ and $|b\rangle$, we would have
to impose the corresponding boundary conditions on the real axis,
at $t=-\infty$ and
$t=+\infty$. Shifting the points at infinity into the complex plane
would then require an adjustment in the boundary values of the fields,
leading to complex solutions.
A complete system of states, on the other hand,
can be inserted anywhere in the complex time plane, so by considering
the total probability (\ref{W1}) we allow ourselves to restrict attention
to real solutions.\footnote{
The question remains, which we did not attempt to resolve in this
paper, whether additional singularities that a solution can have in the
complex time plane are crossed when we deform the contour into the shape
shown in Fig.1. One can verify, however, that no such singularities
have to be crossed in the case of a single symmetric quartic oscillator
where singular solutions can be easily found analytically.}

The derivation of the functional-integral representation for
the probability (\ref{W1}) using the representations for
the $S$-matrix and the projector ${\cal P}_E$ given in \cite{period}
is now completely straightforward and we omit the lengthy intermediate
expressions.
The leading semiclassical result for the transition probability has the
form
\be
W(E_1,E_2) \sim \exp\left( E_1 T_1 - E_2 T_2 - 2 S \right)
\equiv \exp(-Q) \; ,
\label{form}
\ee
where $S$ is the net euclidean action acquired on the euclidean
and anti-euclidean parts of the contour of Fig.1.
The action $S$ corresponds to a solution that
at $t=iT_1/2$ starts from a turning point with energy $E_1$,
follows euclidean evolution to a singularity at $t=0$,
then follows anti-euclidean evolution to a turning point with
energy $E_2$ at $t=iT_2/2$. The actions corresponding to
the euclidean and anti-euclidean evolutions separately are infinite
but they come with opposite signs and when added together give
a finite net amount $S$.

The minkowskian parts of the contour
do not contribute to \eq{form}. Therefore, in applications of \eq{form}
we can forget we ever needed to couple our system to the photon
"bath" because it is always possible to choose the coupling to photons
so small that it does not influence the euclidean and anti-euclidean
parts of the evolution.

As we will see in the next section,
there are infinitely many singular solutions of the form described
above, corresponding to infinitely many possibilities to choose
the initial turning point at $t=iT_1/2$. However, there is a preferred
one, for which $Q$ in (\ref{form}) attains the smallest
possible value at given $E_1$ and $E_2$. This solution saturates
the semiclassical transition probability and determines the
states at $t=-\infty + iT_{1}/2$ and $t=+\infty + iT_2/2$. It also
determines $T_1$ and $T_2$ for given $E_1$ and $E_2$.

An analogous calculation in field theory will
have to impose a momentum cutoff that specifies what modes are considered
"soft". This is not an ambiguity of the method because such a cutoff
has to be built in any procedure of identifying "soft"
particle production in experiment. One might think then that the transition
probability is saturated by a solution for which the final state $|b\rangle$
has particles only in a few "hardest" of the remaining "soft" modes, because
this allows for smaller multiplicity. However, it is not so clear
what the actual answer is: the tendency towards smaller multiplicity
has to compete with the tendency to have a spatially localized coherent field
configuration, which enhances the probability by achieving stronger
non-linearity. Our computations for two oscillators described in Sect.3 show
that the tendency towards stronger non-linearity is indeed rather effective in
distributing energy between "harder" and "softer" modes in the final state.

\section{Numerical results}
We now turn to numerical results for the case of two interacting quartic
oscillators. The hamiltonian is
\be
H= \half {\dot x}^2 + \half {\dot y}^2 + V(x,y) \; ,
\label{ham}
\ee
where the potential $V(x,y)$ is
\be
V(x,y) = \half x^2 + \half \omega^2 y^2 + \frac{\lambda}{4}
\left( x^4 + 6 x^2 y^2 + y^4 \right) \; .
\label{pot}
\ee
The particular choice of the ratios of couplings in (\ref{pot}) is
motivated by the property that the system (\ref{ham})-(\ref{pot})
is equivalent to the $\lambda\phi^4$ theory on a space "lattice" consisting
of two sites. The euclidean and anti-euclidean parts of classical
evolution can be viewed as mechanical motions in the potential
equal to $-V(x,y)$.

Instead of the coordinates  $x$ and $y$ it is convenient to use two
other variables -- the polar angle $\phi$ in the $(x,y)$ plane
and the value of the potential $V$.
The euclidean part of a singular solution starts with zero
velocity at some angle $\phi_i$ and $V=E_1\neq 0$,
and the anti-euclidean part stops at angle $\phi_f$ and
$V=E_2>E_1$. The main question is how we match these two parts at
the singularity.

To formulate the matching condition, consider
the following limiting procedure. Let us choose some large value
of the potential $\L\gg E_1,E_2$. The euclidean trajectory starting
with zero velocity
at $\phi=\phi_i$, $V=E_1$ will reach, in some time $T_1(\L)$,
the value $V=\L$ at some angle $\phi=\phi_{\L}$.
The euclidean action accumulated on this trajectory is some $S_1(\L)$.
We now look for the trajectory that starts with zero velocity at
some angle $\phi_f(\L)$ and $V=E_2$ and reaches, in some time
$T_2(\L)$, exactly the same point $V=\L$, $\phi=\phi_{\L}$ as the
first one.
Mechanical intuition suggests that such trajectory always exists
and our numerical results also show so.
The corresponding euclidean action is some $S_2(\L)$.
Due to anharmonicity, both trajectories reach infinite values of $V$
in finite times, in other words, limits
\be
T_{1,2} = \lim_{\L\to\infty} T_{1,2}(\L)
\label{T}
\ee
always exist. A far more subtle property is the existence of
limits
\begin{eqnarray}
S &= & \lim_{\L\to\infty} \left( S_1(\L) - S_2(\L) \right) \; ,
\label{S} \\
\phi_f & = & \lim_{\L\to\infty} \phi_f(\L) \; .
\label{phif}
\end{eqnarray}
If these limits exist, they define the singular solution
corresponding to given $E_1$, $E_2$ and a given initial configuration
$\phi_i$. We then scan among all initial configurations of energy $E_1$
and find the one that leads to the smallest suppression factor in
(\ref{form}). To summarize, the euclidean and anti-euclidean parts
of a solution are
matched at some large value of the potential energy $\L$ and then
limit $\L\to\infty$ is taken in the hope that the quantities
(\ref{S})-(\ref{phif}) remain stable. In what follows, we present
numerical evidence in favor of the existence of the limits
(\ref{S})-(\ref{phif}).

Numerically, matching at finite $\L$ is simplified by the following
property of euclidean trajectories, which we found empirically.
Suppose there are three trajectories starting with $V=E$ and zero
velocity at angles $\phi_1$, $\phi_2$ and $\phi_3$ and passing
through $V=\L$ at angles $\Phi_1$, $\Phi_2$ and $\Phi_3$, respectively.
The property is that if $\phi_1<\phi_2<\phi_3$, then
$\Phi_1<\Phi_2<\Phi_3$. This "monotony" allows us to use the familiar root
finding methods of bisections and secants, to find
the trajectory that starts at $V=E_2$ and passes $V=\L$ at a specified
value of the angle.

For our computations we chose the following values of the parameters:
$\omega=1.5$, $\lambda=0.1$, $E_1=0.0001$. Such a small value of $E_1$
means that the system is excited out of its ground state, which is the case
of main interest in many applications.
We consider a set of different values of the final state energy $E_2$
in order to obtain the dependence of the suppression factor (\ref{form})
on the excitation energy.

In a numerical computation, it is of course impossible to make
$\L$ arbitrarily large.
Let us estimate the error introduced by the finiteness
of $\L$.
In the Landau-Lifshitz formula for a quartic oscillator, with
$V(x)=x^2/2+g x^4/4$, the absolute value of the error in
$Q/2$ introduced by the finiteness of $\L$ is
\be
\sqrt{2}\int_{x_{\L}}^{\infty} \left(
(V(x)-E_1)^{1/2} - (V(x)-E_2)^{1/2} \right) dx =
\frac{E_2-E_1}{(g\L)^{1/4}} + {\cal O} (\L^{-3/4}) \; .
\label{erl}
\ee
Guided by \eq{erl}, we estimate the same error for our system (assuming
the limits (\ref{S})-(\ref{phif}) exist) as
\be
\frac{E_2-E_1}{(\le\L)^{1/4}} \leq \frac{E_2-E_1}{(\lambda\L)^{1/4}} \; ,
\label{err}
\ee
where $\le$ is an effective coupling constant for a given region of angles,
and the inequality follows from the fact that the smallest anharmonicity
is along the $x=0$ and $y=0$ axes, and this corresponds to $\le=\lambda$.

If we want the error (\ref{err}) to be smaller than some $\epsilon$,
it is sufficient to take $\L$ equal to or larger than
\be
\L_{\epsilon}=
\frac{1}{\lambda} \left( \frac{E_2-E_1}{\epsilon} \right)^4 \; .
\label{Leps}
\ee
Large values of $\L$ dictate the need for high accuracy in solving
the euclidean evolution equations and matching euclidean and anti-euclidean
parts of a solution, in order to insure that numerical
errors coming from those sources do not add up to quantities larger
than $\epsilon$.
Suppose the numerical error we make in solutions for $x(t)$ and $y(t)$
is some $\delta$.
The abbreviated actions $S_{1,2}(\L)-E_{1,2} T_{1,2}(\L)/2$
of the euclidean and anti-euclidean parts of a solution are found according to
\be
S_{1,2}(\L)= 2\int_0^{T_{1,2}(\L)/2}
\{ V[x_{1,2}(t),y_{1,2}(t)]-E_{1,2} \} dt \; .
\label{int}
\ee
Apart from solutions staying on the $x=0$ and $y=0$ axes, for which our
system reduces to that of one degree of freedom, every euclidean solution
eventually reaches a region where
the absolute values of both $x$ and $y$ are large.
This is the region where the main error in $S_{1,2}(\L)$ comes
from. In such a region, the system (\ref{ham}) with our
particular choice of the potential, \eq{pot}, is effectively separable
in terms of variables $\xi=(x+y)/\sqrt{2}$ and $\eta=(x-y)/\sqrt{2}$,
so the main error in $S_{1,2}(\L)$ is the sum of errors in
corresponding abbreviated actions. The errors in $\xi$ and $\eta$ are
not larger than $\sqrt{2}\delta$. Then, for example, the error in the
potential for $\xi$, $V_{\xi}$, is
$\sqrt{2}\delta \partial{V_{\xi}}/\partial{\xi}$  and the error in the
abbreviated action for $\xi$ is estimated as
\[
2\sqrt{2}\delta  \int^{T/2}
\frac{d V_{\xi}}{d \xi} dt =
2\delta \int^{\xi_\L}
\frac{(d V_{\xi}/d \xi)~d\xi}{(V_{\xi}-E_{\xi})^{1/2}}
\]
\be
= 4\delta\sqrt{V_{\xi}(\xi_\L)} +
{\cal O}(\delta/\sqrt{V_{\xi}(\xi_\L)}) \; .
\label{dint}
\ee
(Strictly speaking, some numerical error comes also from determination of
$T_{1,2}$ but that can be easily made much smaller than (\ref{dint}).)
Even though both $V_{\xi}(\xi_\L)$ and $V_{\eta}(\eta_\L)$ are large,
only one of them is of order (and hence essentially equal to) $\L$.
So, the numerical error in each of the abbreviated actions (\ref{int}),
produced by the error $\delta$ in the solution, is estimated as
$4\delta\sqrt{\L}$. If we want the corresponding error in $Q/2$ to be much
smaller than $\epsilon$, we have to make $\delta$ much smaller than
\be
\delta_{\epsilon} (\L) = \frac{\epsilon}{8\sqrt{\L}} \; .
\label{deps}
\ee
For the smallest admissible $\L$, \eq{Leps}, this becomes
\be
\delta_{\epsilon} (\L_{\epsilon})
= \frac{\epsilon^3 \sqrt{\lambda}}{8 (E_2-E_1)^2} \; .
\label{ddeps}
\ee
We see that even moderate improvements in the desired accuracy $\epsilon$
of $Q/2$ require considerable improvements in the accuracy of the solution.
The required accuracy of the solution is also higher for transitions
between states that are more separated in energy.

Our computations were done with $\epsilon=0.1$.
The equations of motion were solved and euclidean and anti-euclidean
parts were matched to accuracy more than an order of magnitude better
than (\ref{deps}).
Table 1 illustrates the existence of the limits (\ref{S})-(\ref{phif})
for $E_2=15$ and $\phi_i=\pi/6$. Note that for $\L\geq \L_{\epsilon}$,
the value of $Q/2$ does not change with $\L$ within the accuracy $\epsilon$,
in correspondence with our estimates.
\begin{table}
\begin{tabular}{|l|rrrrr|}
\hline
$\L$ & $0.5\times \L_{\epsilon}$ & $\L_{\epsilon}$ &
$2.5\times \L_{\epsilon}$ &
$5\times \L_{\epsilon}$ & $10\times \L_{\epsilon}$   \\
\hline
$Q/2$ & 23.05 & 23.06  & 23.08 & 23.09 & 23.10 \\
\hline
$\phi_f$ & 1.40979841 & 1.40979916 & 1.40979996 & 1.40980045 & 1.40980085 \\
\hline
\end{tabular}
\caption{Values of half of the exponent $Q$ of eq.(3) and the final state
angle $\phi_f$ for $E_2=15$, initial state angle $\phi_i=\pi/6$ and different
values of the matching energy $\L$. $\L_{\epsilon}=5.06\times 10^9$.}
\end{table}
We remind that each value of $Q/2$ in Table 1 is obtained as a difference
between large (divergent at $\L\to\infty$) abbreviated actions
$S_1-E_1 T_1/2$
and $S_2-E_2 T_2/2$ acquired on the euclidean and anti-euclidean parts of
the solution. For example, for $\L=5\times V_{\epsilon}$, the values of
these abbreviated actions are of order $6\times 10^7$.

Next we consider values of the exponent $Q$ for fixed
$E_2$ and different values of the initial angle $\phi_i$.
Results for $E_2=15$ are presented in Table 2.\footnote{
These results were actually obtained by finding
$\phi_i$ for a set of $\phi_f$ equally spaced between $0$ and $\pi/2$.
Equal spacing of $\phi_f$, rather than of $\phi_i$, happens
to give more
points in the most interesting region of angles where $Q$ is close to its
minimum.}
\begin{table}
\begin{tabular}{|l|rrrrrrrrrr|}
\hline
$\phi_i$ & $\sim 10^{-19}$ & 0.008 & 0.017 & 0.028 & 0.041 &
0.058 & 0.083 & 0.129 & 0.231 & 0.537 \\ \hline
$\phi_f$ & 0 & $\pi/20$ & $\pi/10$ & $3\pi/20$ & $\pi/5$ &
$\pi/4$ & $3\pi/10$ & $7\pi/20$ & $2\pi/5$ & $9\pi/20$ \\ \hline
$Q/2$ & 25.2 & 23.0 & 21.9 & 21.3 & 21.1 & 21.0 & 21.2 &
21.6 & 22.2 & 23.1 \\ \hline
\end{tabular}
\caption{Several values of the initial angle and the corresponding
values of the final angle and $Q/2$ for $E_2=15$. Estimated error
in $Q/2$ is less than 0.1.}
\end{table}
We see that $Q$ has a minimum near $\phi_f=\pi/4$.
The corresponding value $Q/2=21.0$ determines the suppression factor
(\ref{form}) for $E_2=15$.

Note that the preferred final state is not purely the "harder" $y$-component
but a state where $x$ and $y$ are comparable, which achieves stronger
non-linearity.  This tendency persists to larger values of $\omega$.
For example, at $\omega=3$, $E_2=10$, very few $y$-quanta need to be
produced but the system prefers $\phi_f$ near $3\pi/10$ rather than
$\phi_f=\pi/2$.

The minimum in $Q/2$ becomes more
pronounced as the final state energy $E_2$ increases.
Table 3 gives such minimal values of $Q/2$ for various values of $E_2$.
\begin{table}
\begin{tabular}{|l|rrrr|}
\hline
$E_2$ & 10 & 15 & 25 & 50  \\ \hline
$Q/2$ & 15.4 & 21.0 & 30.9 & 51.8  \\
\hline
\end{tabular}
\caption{Values of $Q/2$ determining suppression factors
(3) for initial state energy $E_1=0.0001$ and various final
state energies $E_2$. Estimated error in $Q/2$ is less than 0.1.}
\end{table}
The suppression increases with the excitation energy, as may be
intuitively expected.

\section{Tunneling probabilities}
In the previous sections, we considered transition probabilities,
for which we summed over all initial and final states
of given energies (or, more precisely, states with energies
distributed in a small interval near given ones -- microcanonical
states). For potentials with two or more local minima, at energies less
than the height of the barrier, we may be interested separately in the
probability of a tunneling process,
when a system originally distributed near energy
$E_1$ in one well gets excited to the state near energy $E_2$
in another. If $E_1=E_2$, tunneling probabilities for microcanonical
initial states can be calculated
using regular euclidean solutions -- periodic instantons \cite{period}.
When $E_1\neq E_2$, however, we again have to use singular solutions.

In the case of a single degree of freedom, the relevant singular solution
is easily found (see for example ref.\cite{DP}). Let us assume $E_1<E_2$.
The periodic instanton corresponding
to energy $E_2$ has two turning points $x_2$ and $x_3$, one of
them, $x_2$, in the well where the system is originally located.
If we use $x_2$, with zero velocity,
as an initial condition for {\em minkowskian} evolution,
the resulting (real) minkowskian trajectory will stop again at a different
point $x_1$ in the same well. This new
turning point $x_1$, which still has energy $E_2$, can now be used as the
final point of the anti-euclidean segment of a singular euclidean solution
whose euclidean segment starts at some $x_0$ with energy $E_1$.
The whole trajectory from $x_0$ to $x_3$ comprises the singular solution
that describes the tunneling process. The corresponding suppression exponent
is the sum of the exponent $Q$, \eq{form}, acquired from $x_0$
to $x_1$, and the suppression exponent associated with the periodic instanton.

The stopping of the minkowskian trajectory at $x_1$ is a special
feature of systems with one degree of freedom. When there are several degrees
of freedom, velocity has several components which in general will never vanish
simultaneously. Further, there seems to be no reason why
they should vanish simultaneously (except at the initial point)
for the minkowskian trajectory starting at the turning point of
a periodic instanton. However, a numerical "experiment" shows that
in a double-well system of two interacting quartic oscillators (see below)
this in fact {\em nearly} happens.
In other words, the minkowskian solution starting at the turning point of a
periodic instanton is approximately periodic.

Because the periodicity of the minkowskian solution is not exact, the
method described above for the
case of one degree of freedom cannot be literally applied to
several degrees of freedom. Most likely, to generalize this method to
several degrees of
freedom, one has to consider complex singular solutions.
The approximate periodicity is still of value, however, because
it allows us to construct a purely real {\em approximate} singular
solution describing a tunneling transition and use it
to {\em estimate} the probability of that transition.

Specifically, consider the hamiltonian (\ref{ham}) with
the following double-well potential,
\be
V(x,y) = -\half x^2 + \half \omega^2 y^2 + \frac{\lambda}{4}
\left( x^4 + 6 x^2 y^2 + y^4 \right) - \alpha y \; .
\label{potd}
\ee
The term $-\alpha y$ was added to avoid the trivial situation when stopping
occurs simply because $y=0$ on the whole trajectory.
Below we present results for $\omega=1.5$, $\lambda=0.01$, $\alpha=1$ and
$E_2=-15$ (minima of the potential (\ref{potd}) have negative energy).
Similar results were obtained for other values of the parameters
and for other quartic double-well potentials.
The approximate stopping property is less pronounced for potentials of
higher order.

We found it convenient to characterize configurations in the left well
by their polar angles $\phi$ in the polar system in which the origin is
placed at $x=-1/\lambda^{1/2}$, $y=0$. (This would be the location of the left
minimum of the potential at $\alpha=0$.)
The turning point of the periodic instanton for the quoted values of the
parameters is $\phi=0.0817$. The absolute value of velocity
$v=({\dot x}^2 + {\dot y}^2)^{1/2}$ on the minkowskian trajectory starting
at this point with zero velocity is plotted as a function of time in Fig.2a.
We see that the absolute value of the velocity almost reaches zero.
Numerically, the minimum absolute value of the velocity (not counting
zero value at $t=0$) in this case is about $3\times 10^{-4}$, which is
at least hundred times smaller than its "natural" value.  For comparison,
Fig.2b shows the same quantity for the minkowskian trajectory starting with
zero velocity at arbitrarily chosen point $\phi=0.01$. Here the absolute
value of velocity does not approach zero (except at zero time) and is
clearly aperiodic.
\begin{figure}
	\def\epsfsize	#1#2{0.51#1}
	\epsfbox [100 250 450 480] {plota.ps}
        \epsfbox [100 250 450 600] {plotb.ps}
\caption{
(a) Absolute value of velocity on the minkowskian trajectory
starting with zero velocity at a turning point of a periodic instanton.
(b) The same for an arbitrarily chosen initial point.}
\end{figure}

We can use the approximate stopping point of a minkowskian solution as
the final state $\phi_f$ for a real singular euclidean solution similar to
those found in Sect.3.
Because the euclidean solution reaches exactly zero velocity at this point
while the minkowskian solution does so only approximately, the trajectory
comprised of the two will not be an exact solution. Still, it is useful for
estimates of tunneling probabilities in our system. We thus describe
the tunneling process as consisting of
two stages: first, the system is excited from energy $E_1$ to energy $E_2$
while staying in the same well, then it tunnels at energy $E_2$.
The tunneling probability is then estimated as
\be
W_{\mbox{tun}}(E_1,E_2) \sim \exp(-Q + E_2 T' - S' )
\equiv \exp(-Q_{\mbox{tun}}) \; ,
\label{tun}
\ee
where $Q$ is associated with the singular euclidean solution in the same way
as in \eq{form}, $S'$ is the action of the periodic instanton and $T'$
is its period. Table 4 shows values of $Q_{\mbox{tun}}/2$ for the potential
(\ref{potd}) with the same values of $\omega$, $\lambda$ and $\alpha$ as above,
$E_1=-24$ and various values of $E_2$.
\begin{table}
\begin{tabular}{|l|rrrrrr|}
\hline
$E_2$ & -24 & -15 & -10 & -5 & -1 & -0.25 \\ \hline
$Q_{\mbox{tun}}/2$ & 87.6 & 68.5 & 59.4 & 50.8 & 44.3 & 43.0 \\
\hline
\end{tabular}
\caption{Values of $Q_{\mbox{tun}}/2$ determining estimated suppression
factors (10) for the potential (9), initial state energy $E_1=-24$ and
various final
state energies $E_2$. Estimated numerical error in
$Q_{\mbox{tun}}/2$ is less than 0.1.}
\end{table}
For reference, the energy of the saddle point separating two minima is
$E_0=-0.222$.
We see that the suppression decreases as this energy is approached, though
it does not go away completely.

\section{Conclusion}
Our aim in this work was to introduce the boundary problem for
singular euclidean solutions and to gain evidence for its correctness
by doing numerical calculations for a system of two interacting
non-linear oscillators. For two oscillators with a quartic double-well
potential, we also presented numerical evidence that the
{\em minkowskian} trajectories starting with zero velocity from turning
points of periodic instantons (which are periodic {\em euclidean} solutions)
have approximate stopping points or, equivalently, are approximately
periodic.
This property allowed us to construct real approximate singular solutions
describing tunneling transitions and use them to estimate
the probabilities of such transitions.

One may try to extend the latter approach to other systems, including
field theories. The accuracy to which the approximate stopping property
holds will differ for different systems, and so will the accuracy of
estimates based on it. However, within that accuracy, this
approach gives an in principle straightforward method, essentially identical
to the method used for a single degree of freedom, of numerical estimates
of tunneling excitation rates in systems with many degrees of freedom.

The author is grateful to V. Petrov, V. Rubakov and P. Tinyakov for
useful discussions, and to the Aspen Center for Physics, where some of
these discussions took place, for hospitality. Part of this work was done
while the author was at the University of California, Los Angeles as
a Julian Schwinger postdoctoral fellow.

\end{document}